\documentclass[reprint,aps,prc,twocolumn,superscriptaddress,longbibliography,floatfix,10pt]{revtex4-1}

\usepackage{bm}
\usepackage{dcolumn}
\usepackage{amsthm}
\usepackage{amssymb}
\usepackage{amsmath}
\usepackage{graphicx} 
\usepackage{bm}
\usepackage{xcolor}
\usepackage{fix-cm}
\usepackage{mathptmx} 
\usepackage[T1]{fontenc}
\usepackage[colorlinks,allcolors=blue]{hyperref}
\makeatletter
\def\NAT@def@citea{\def\@citea{\NAT@separator}}
\makeatother
\renewcommand{\>}{\rangle}
\newcommand{\<}{\langle}

\renewcommand{\vr}{{\bf r}}

\newcommand{\mH}{{\mathcal{H}}}

\renewcommand{\d}{{\mbox d}}

\newcommand{\sdf}{\,\,}

\begin{document}   


\title{Absence of hindrance in microscopic \textsuperscript{12}C+\textsuperscript{12}C fusion study}


\author{K. Godbey}
\email[]{kyle.s.godbey@vanderbilt.edu}
\affiliation{Department of Physics and Astronomy, Vanderbilt University, Nashville, Tennessee 37235, USA}

\author{C. Simenel}
\email[]{cedric.simenel@anu.edu.au}
\affiliation{Department of Theoretical Physics and Department of Nuclear Physics, Research School of Physics and Engineering, The Australian National University,
	Canberra ACT 2601, Australia}

\author{A. S. Umar}
\email[]{sait.a.umar@vanderbilt.edu}
\affiliation{Department of Physics and Astronomy, Vanderbilt University, Nashville, Tennessee 37235, USA}

\date{\today}

\begin{abstract}
\edef\oldrightskip{\the\rightskip}
\begin{description}
	\rightskip\oldrightskip\relax
	\setlength{\parskip}{0pt}
	\item[Background] Studies of low-energy fusion of light nuclei are important in astrophysical modeling, with small variations in reaction rates having a large impact on nucleosynthesis yields. Due to the lack of experimental data at astrophysical energies, extrapolation and microscopic methods are needed to model fusion probabilities.
	\item[Purpose] To investigate deep sub-barrier \textsuperscript{12}C+\textsuperscript{12}C fusion cross sections and establish trends for the $S$ factor.
	\item[Method] Microscopic methods based on static Hartree-Fock (HF) and time-dependent Hartree-Fock (TDHF) mean-field theory are used to obtain \textsuperscript{12}C+\textsuperscript{12}C ion-ion fusion potentials. Fusion cross sections and astrophysical $S$ factors are then calculated using the incoming wave boundary condition (IWBC) method.
	\item[Results] Both density-constrained frozen Hartree-Fock (DCFHF) and density-constrained TDHF (DC-TDHF) predict a rising $S$ factor at low energies, with DC-TDHF predicting a slight damping in the deep sub-barrier region ($\approx1$~MeV). Comparison between  DC-TDHF calculations and maximum experimental cross-sections in the resonance peaks are good. However the discrepancy in experimental low energy results inhibits interpretation of the trend.
	\item[Conclusions] Using the fully microscopic DCFHF and DC-TDHF methods, no $S$ factor maximum is observed in the \textsuperscript{12}C+\textsuperscript{12}C fusion reaction. In addition, no extreme sub-barrier hindrance is predicted at low energies. The development of a microscopic theory of fusion including resonance effects, as well as further experiments at lower energies must be done before the deep sub-barrier behavior of the reaction can be established. 
\end{description}
\end{abstract}


\maketitle

\section{Introduction}
\label{sec:intro}
The study of sub-barrier fusion between light nuclei is of paramount importance to astrophysical applications ranging from element formation to accreting neutron star superbursts.
The \textsuperscript{12}C+\textsuperscript{12}C fusion reaction in particular stands as a fundamental part of these processes.
For example, large type~I x-ray bursts (known as superbursts) are suspected to arise from unstable thermonuclear carbon burning in neutron stars accreting material from their partner in binary systems~\cite{cumming2001,strohmayer2002}.
Another example of the importance of this reaction is that it is  a vital step on the path to heavier nuclei in the final stages of the helium burning process~\cite{hoyle1954}.
Reaction rates of \textsuperscript{12}C+\textsuperscript{12}C fusion influence both the immediate formation of A~$\geq 20$ systems and the creation of larger elements later in the nucleosynthesis processes.

Despite their importance, the \textsuperscript{12}C+\textsuperscript{12}C fusion reaction rates at deep sub-barrier energies are poorly known. 
Experimentally, this is of course due to the extremely low tunneling probabilities at astrophysical energies 
and the relatively high backgrounds from unwanted reactions~\cite{spillane2007,zickefoose2011}, 
motivating for the development of new particle-$\gamma$ coincidence techniques to reduce this background \cite{courtin2017,jiang2018}. 
Resonances in the fusion excitation functions could also be present at deep sub-barrier energies \cite{cooper2009,tumino2018} 
and could potentially modify capture rates by orders of magnitudes in this energy range. 
As a result, with current experimental uncertainties, 
the bounds of the C-burning reaction rate curves result in a wide range of $s$-process and $p$-process abundances~\cite{pignatari2012}.
New measurements on nearby systems  like \textsuperscript{13}C+\textsuperscript{12}C \cite{zhang2018}
have also been performed to establish bounds for the low energy behavior of \textsuperscript{12}C+\textsuperscript{12}C, 
as the large resonances and oscillations in the sub-barrier region are absent from the asymmetric systems. 

On the theory side, predictions of fusion cross-sections at such low energies vary by orders of magnitude and are thus not able to confidently guide experiments. 
To assist astrophysical models, attempts have been made to extrapolate \textsuperscript{12}C+\textsuperscript{12}C fusion cross sections to lower energies of astrophysical interest. 
One common method of extrapolation involves fitting a phenomenological model describing $S$ factors or cross sections to experimental results and extrapolating the values to lower energies~\cite{fowler1975,jiang2007}.
The behavior of these models depend on the chosen formula for the fit, resulting in radically different outcomes.
Furthermore, these models 
do not produce the required conditions for superbursts to occur in binary systems~\cite{cumming2006}, indicating a need for other reactions to provide additional heat, or the presence of resonances~\cite{cooper2009} not accounted for in these extrapolations. 



Theoretical developments to improve the description of sub-barrier fusion are based on various strategies.
A common approach is to utilize a phenomenological nucleus-nucleus potential and calculating quantum tunneling probabilities for transmission through this potential.
One such method utilizes the S\~ao Paulo potential, a model that is tuned for astrophysical applications. 
This model allows for the inclusion of dynamical polarization of the incoming nuclei via an energy dependence of the potential~\cite{gasques2004,gasques2005,gasques2007}.
Alternatively, polarization effects can be included via the coupled-channels method based on a bare (energy independent) nucleus-nucleus potential, 
e.g., calculated with a modified double-folding technique~\cite{esbensen2011,jiang2013}.
Deformation effects of \textsuperscript{12}C have also been investigated with potential models~\cite{denisov2010c}, microscopic calculations~\cite{heenen1981},
and the time-dependent wave-packet method in an attempt to reproduce the resonances that show in \textsuperscript{12}C+\textsuperscript{12}C fusion~\cite{diaz-torres2018}.

Another strategy is to incorporate dynamical polarization effects via fully microscopic time-dependent approaches 
such as the time-dependent Hartree-Fock (TDHF) mean-field theory~\cite{negele1982,simenel2012,simenel2018}. 
The main motivations for using TDHF as a tool to study ion-ion fusion is that 
{\it (i)} it incorporates unrestricted shape evolution of the collision partners induced by coupled-channel effects \cite{simenel2013b}, and
{\it (ii)} the only parameters of the theory are those of the energy density functional (usually of the Skyrme type~\cite{skyrme1956}, though results using the quark-meson coupling (QMC)~\cite{guichon1988,guichon1996,stone2016} approach are also presented here) 
describing the effective interaction between the nucleons. 
Direct applications of TDHF to light systems have been performed at and above the fusion barrier \cite{bonche1978,lebhertz2012,simenel2013a}. 
A drawback of the approach, however, is that it does not incorporate quantum tunneling of the many-body wave-function 
and thus cannot be used directly to describe sub-barrier fusion. 

To overcome this limitation, we predict the nucleus-nucleus potential from TDHF calculations, and then compute transmission probabilities through this potential. 
The dynamic evolution of the nuclei during the collision leads to an energy dependence of the ion-ion potential as the nuclear densities 
in near-barrier collisions morph much more than in high energy collisions~\cite{washiyama2008,umar2014a,jiang2014}. 
Using TDHF to predict nucleus-nucleus potentials has important advantages such as a proper treatment of the Pauli exclusion principle \cite{simenel2017} 
and the inclusion of coupling to transfer channels \cite{godbey2017}.
Its main limitations, however, are that {\it (i)} the effect of the couplings are only treated in average 
(while in coupled-channels calculations the incoming channels are summed coherently) and 
{\it (ii)} the energy dependence of the potential is only accounted for at above barrier energies.
The second limitation, in particular, is problematic as there is no guarantee that the dynamics of the couplings at the barrier are the same as in deep sub-barrier energies. 
Nevertheless, predictions of sub-barrier fusion cross-sections based on nucleus-nucleus potentials computed from TDHF trajectories just above the barrier 
are usually in good agreement with experiments \cite{umar2012a,umar2009b,keser2012,simenel2013a}.


The aim of the present work is to test predictions of  \textsuperscript{12}C+\textsuperscript{12}C fusion cross-sections 
at deep-sub-barrier energies with potentials derived microscopically from TDHF trajectories at near-barrier energies.
This is achieved with the density-constrained TDHF (DC-TDHF) method. 
To consider the separate static and dynamic effects, the density-constrained frozen Hartree-Fock (DCFHF) method has also been used in this study.
DCFHF, which is a fully microscopic static approach for studying ion-ion potentials, 
allows for more direct comparisons to other static approaches and offers a baseline for differences when dynamics is considered~\cite{simenel2017}. 
Our attention is focused on various factors which could impact the theoretical prediction, in order to test the robustness of the predictions. 
In particular, the effects of numerical approximations, e.g., associated with the grid characteristics, are studied in details. 
Different energy density functionals are also considered.
As the predictions are free of adjustable parameters, our goal is to provide the best possible prediction with existing microscopic tools, 
with the perspective of identifying limitations of the approach and possible future extensions.

The structure of the paper is as follows. A thorough description of both DCFHF and DC-TDHF approaches is presented in Sec.~\ref{sec:form} 
with a general prescription for calculating transmission probabilities presented first. 
Section~\ref{sec:res} then presents the results from both methods as they compare to experimental data 
and recent predictions of the $S$ factor behavior at low energies of astrophysical interest. 
Finally a brief summary of the results and a few closing comments are presented in Sec.~\ref{sec:conclusion}.

\section{Formalism}
\label{sec:form}
In this section we introduce the methods used in computing \textsuperscript{12}C+\textsuperscript{12}C fusion cross sections.
\subsection{Density-constrained frozen Hartree-Fock}
\label{sec:dcfhf}
Following the idea of Brueckner \textit{et al.}~\cite{brueckner1968}, we derive the bare ion-ion potential from an energy density functional (EDF)  $E[\rho]$
written as an integral of an energy density $\mH[\rho(\vr)]$, i.e.,
\begin{equation}
E[\rho]=\int \d\vr \sdf \mH[\rho(\vr)]~.
\end{equation}
The bare potential is obtained by requiring frozen ground-state densities $\rho_{i}$ of each nucleus ($i=1,2$) computed using the Hartree-Fock (HF) mean-field approximation~\cite{hartree1928,fock1930}.
The Skyrme~\cite{skyrme1956} and quark-meson coupling (QMC)~\cite{stone2016} EDFs are used both to calculate the HF ground state, which are found spherical, and to compute the potential.
It accounts for the bulk properties of nuclear matter such as its incompressibility which is crucial at short distances~\cite{brueckner1968,misicu2006,hossain2015}.
Overlaying the densities while neglecting the Pauli exclusion principle between nucleons in different nuclei
leads to the frozen Hartree-Fock (FHF) potential~\cite{washiyama2008,simenel2008,simenel2012}
\begin{equation}
V_{FHF}(R)=\int \d\vr \sdf \mH[\rho_1(\vr)+\rho_2(\vr-R)] - E[\rho_1] -E[\rho_2],
\label{eq:frozen}
\end{equation}
where $R$ is the distance between the centers of the two nuclei.
The resulting FHF potential can then be used to calculate cross sections and related quantities.

The density-constrained FHF (DCFHF) method is the extension of FHF to exactly account for the Pauli exclusion principle between nucleons~\cite{simenel2017}.
This inclusion is obtained by using the same frozen densities $\rho_{i}$ from the FHF initialization as a constraint for a new HF minimization.
This allows the single-particle states to reorganize into a lower energy configuration while maintaining that they are properly antisymmetrized and that the neutron and proton densities remain the same.
The potential is defined similarly to FHF, though with the density-constrained wave functions~\cite{simenel2017}
\begin{equation}
V_{\mathrm{DCFHF}}(R)=\<\Phi_{\mathrm{DC}}(R) | H | \Phi_{\mathrm{DC}}(R) \>-E[\rho_1]-E[\rho_2]\,.
\label{eq:vr}
\end{equation}
The resulting DCFHF potential has the same barrier height as FHF, though the pressure from the Pauli exclusion principle usually forms a pocket inside the fusion barrier.
DCFHF is the static analog to the density-constrained time-dependent Hartree-Fock (DC-TDHF) approach discussed in Sec.~\ref{sec:dctdhf}, and thus is particularly useful to separate static and dynamic effects~\cite{vophuoc2016}.

\begin{figure}
	\includegraphics*[width=8.6cm]{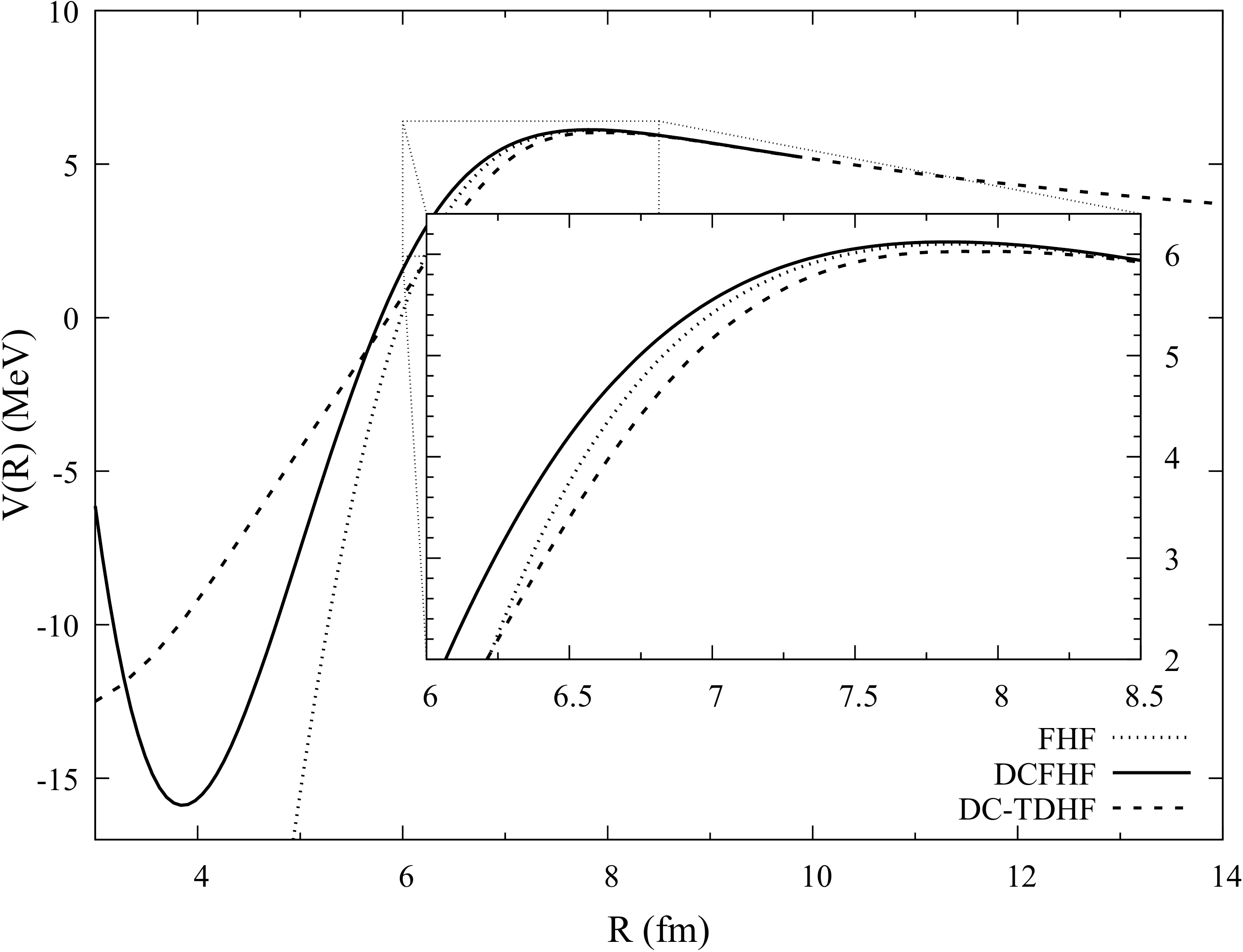}
	\caption{Ion-Ion fusion potentials from DC-TDHF, FHF, and DCFHF using the UNEDF1 force.}
	\label{fig:pots}
\end{figure}
\begin{figure}
	\includegraphics*[width=8.6cm]{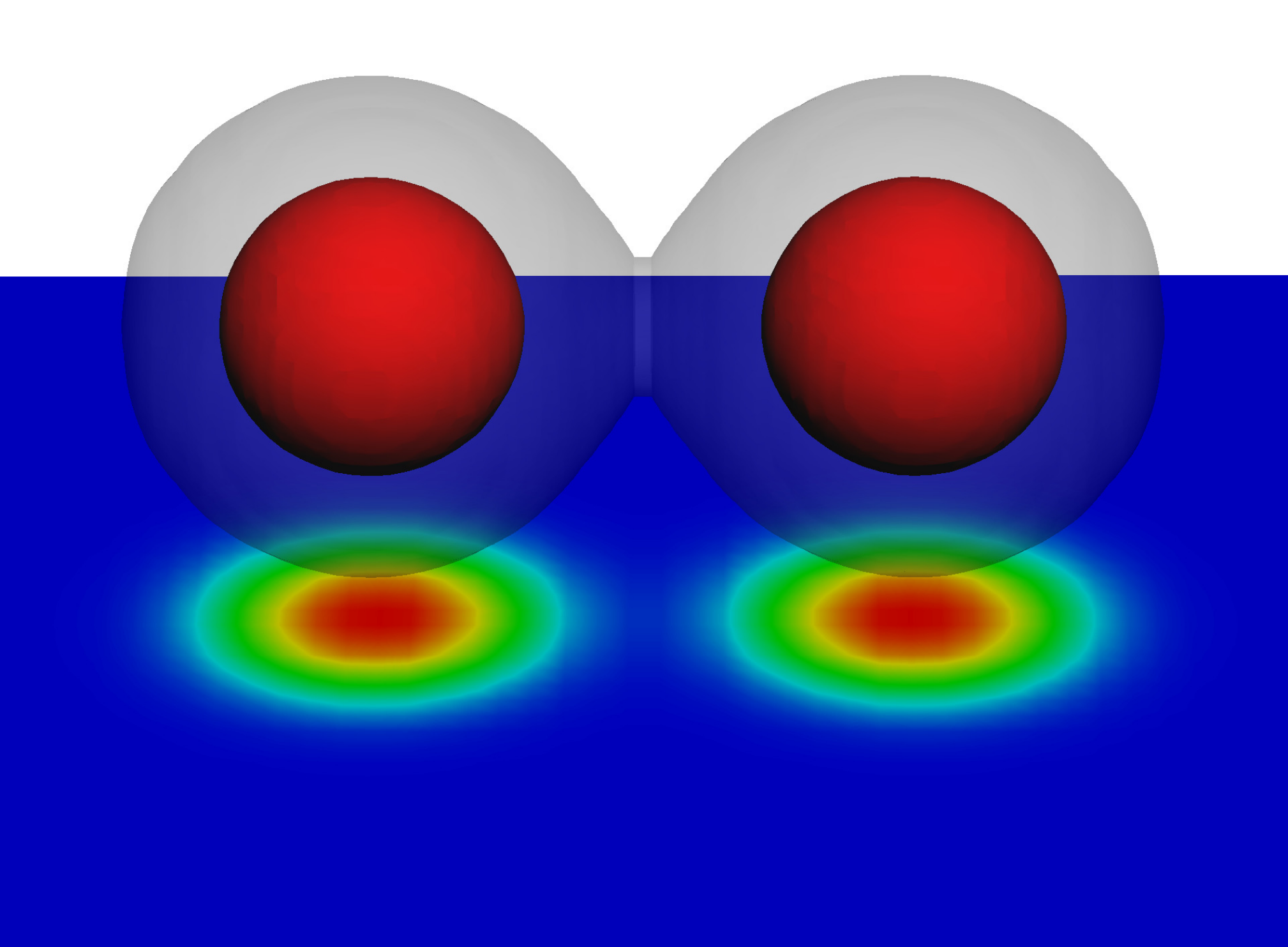}
	\caption{(Color online) 3D contour plot over projected pseudocolor density plot at distance of the barrier peak at $R=8.06$~fm. Solid red contour surface is drawn for $\rho = 0.08$~fm\textsuperscript{-3} and opaque gray shading is drawn for $\rho = 0.008$~fm\textsuperscript{-3} to show the value and location of interacting densities.}
	\label{fig:dens}
\end{figure}
\begin{figure}
	\includegraphics*[width=8.6cm]{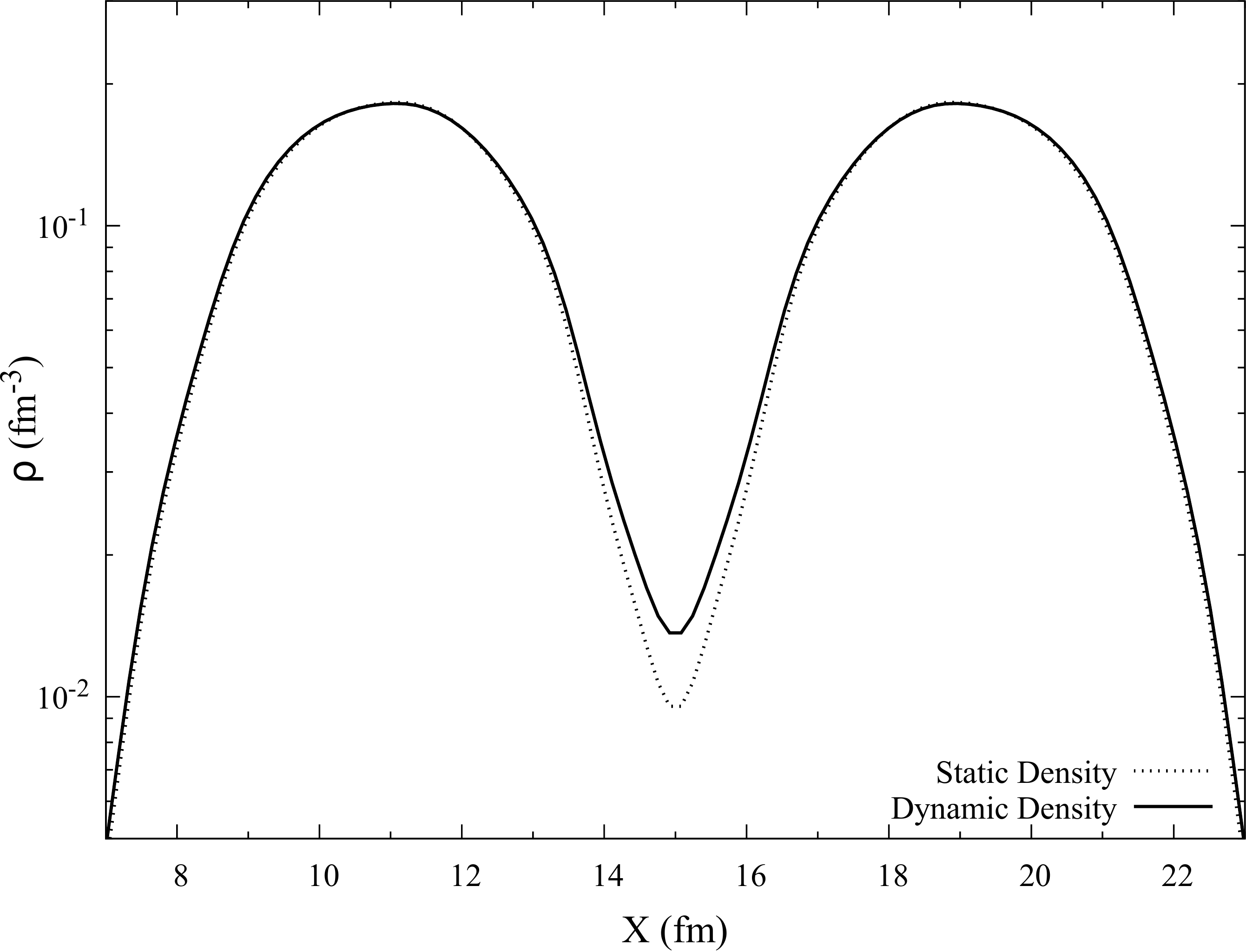}
	\caption{Density profile along the collision axis of the \textsuperscript{12}C + \textsuperscript{12}C system at the barrier peak for densities from the static FHF method and dynamic DC-TDHF method. }
	\label{fig:denstrace}
\end{figure}
\begin{figure*}
	\includegraphics[width=0.49\textwidth]{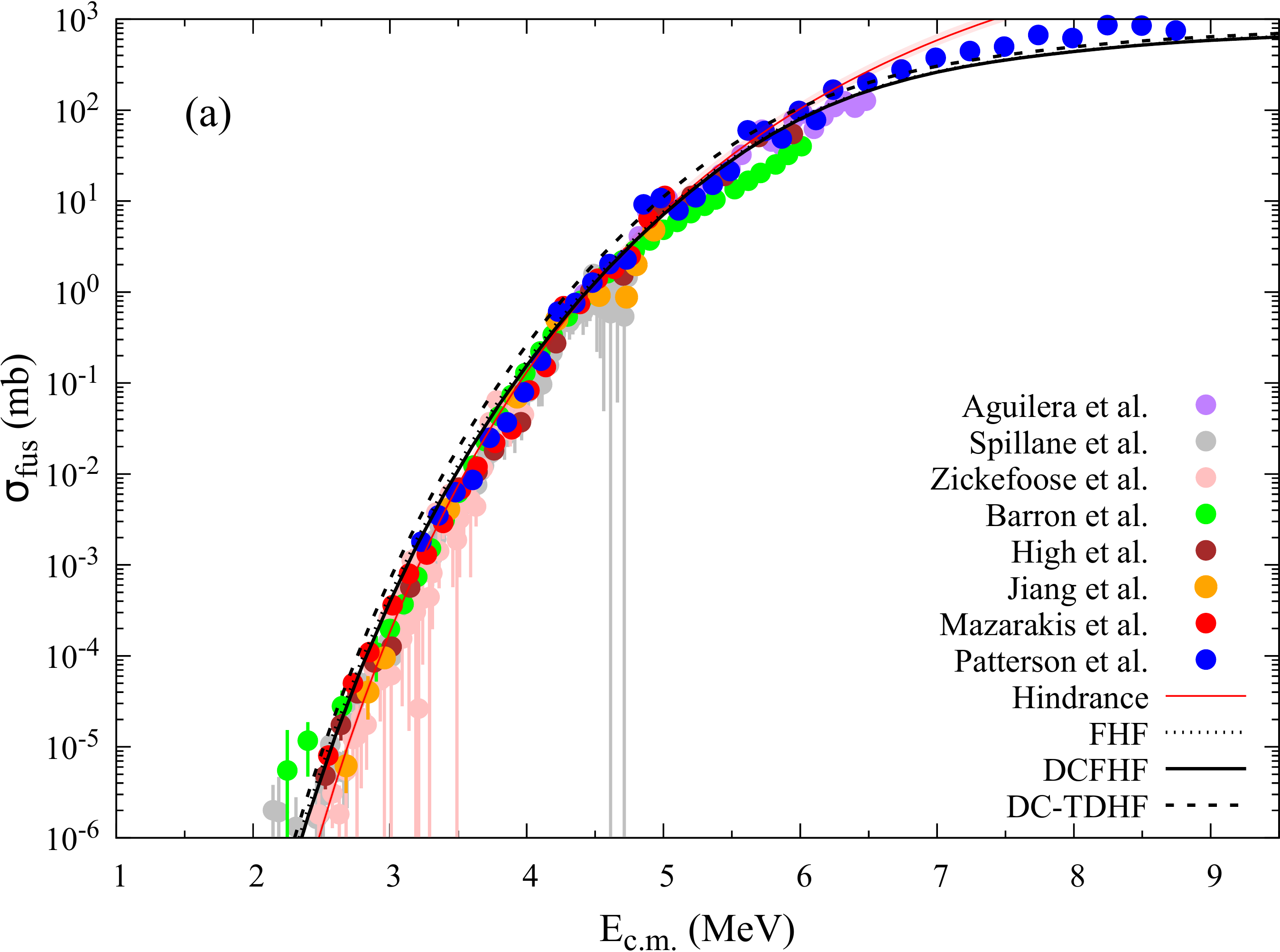}
	\includegraphics[width=0.49\textwidth]{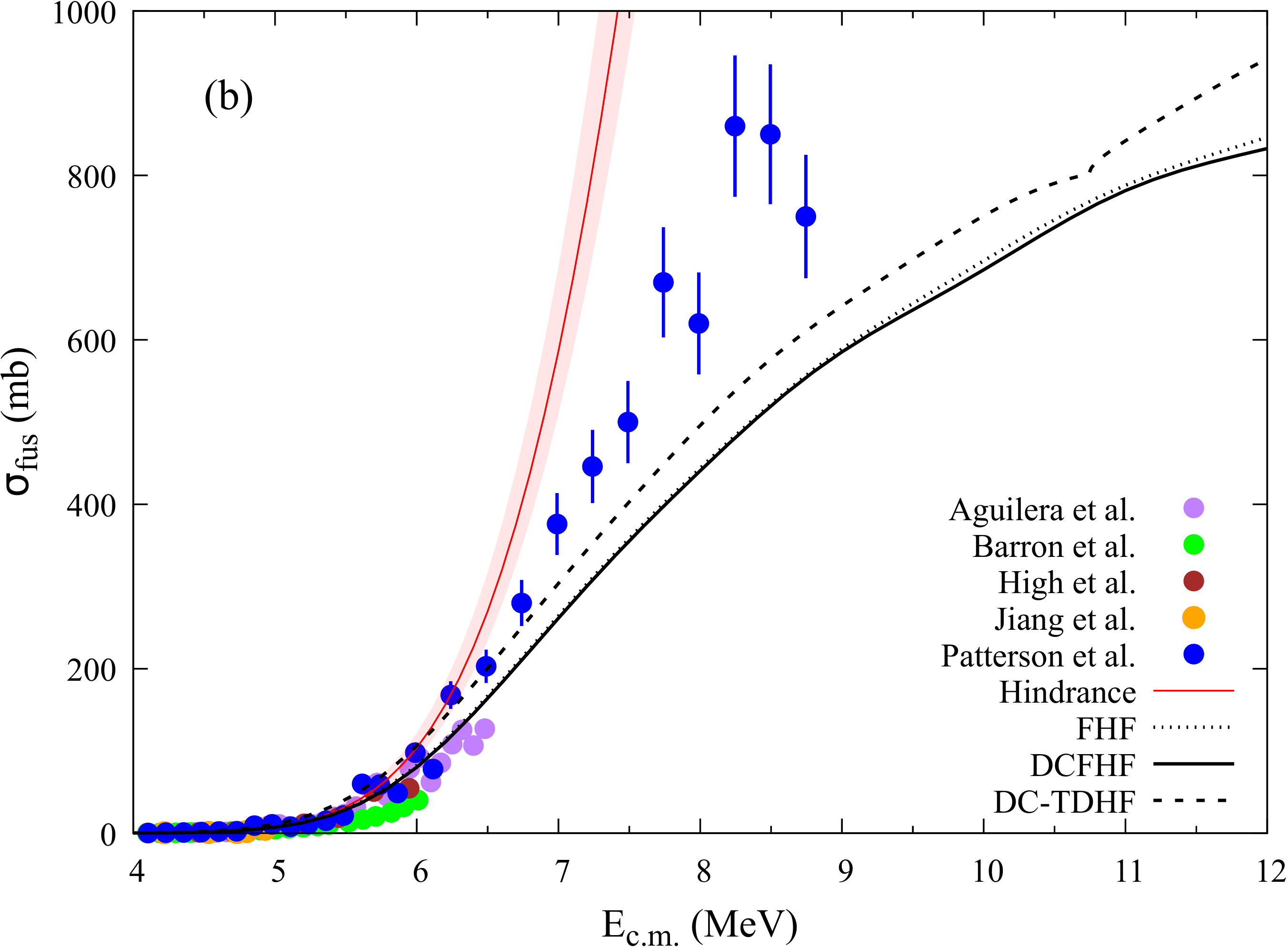}
	\caption{(Color online) Comparison of \textsuperscript{12}C+\textsuperscript{12}C fusion cross sections from DC-TDHF, FHF, and DCFHF using the UNEDF1 force and experimental data from~\cite{aguilera2006,spillane2007,zickefoose2011,barron-palos2006,high1977,jiang2018,mazarakis1973,patterson1969} in (a) logarithmic and (b) linear scales. The hindrance model is from Ref.~\cite{jiang2007}.}
	\label{fig:xsec}
\end{figure*}
\subsection{Density-constrained time-dependent Hartree-Fock}
\label{sec:dctdhf}

In the TDHF approximation the many-body wave function is taken as a single
Slater determinant.
In this limit, the variation of the time-dependent action with respect to the single-particle states, $\phi^{*}_{\lambda}$, yields the most probable time-dependent path
in the multi-dimensional space-time phase space represented as a
set of coupled, nonlinear, self-consistent initial value equations
for the single-particle states
\begin{equation}
h(\{\phi_{\mu}\}) \ \phi_{\lambda} (r,t) = i \hbar \frac{\partial}{\partial t} \phi_{\lambda} (r,t)
\ \ \ \ (\lambda = 1,...,A)\,,
\label{eq:TDHF}
\end{equation}
where $h$ is the HF single-particle Hamiltonian.
These are the fully microscopic time-dependent Hartree-Fock equations.

The DC-TDHF approach~\cite{umar2006b} is then employed to calculate the ion-ion potentials $V_{\mathrm{DC{\hbox{-}}TDHF}}(R)$ directly from TDHF dynamics and has been used to calculate fusion cross sections for a wide range of
reactions~\cite{godbey2017,umar2014a,simenel2013a,umar2012a,umar2006a,oberacker2010,umar2009a,jiang2015a}.
This approach differs from the DCFHF method in that the nuclear density changes in time following the
TDHF evolution.
The main steps of this approach are as follows:
At certain times $t$ or, equivalently, at certain internuclear distances
$R(t)$, a static energy minimization is performed with the same constraint mentioned in Sec.~\ref{sec:dcfhf}, i.e. constraining the proton and neutron densities to be equal to the instantaneous TDHF densities
\begin{equation}
\delta \< \ H - \sum_{q=p,n}\int d\vr \ \lambda_q(\vr) \ \rho^{TDHF}_q(\vr) \ \> = 0\;.
\label{eq:var_dens}
\end{equation}
We refer to the minimized energy as the ``density-constrained energy''
$E_{\mathrm{DC}}(R)$.
The ion-ion interaction potential $V(R)$ is calculated by
subtracting the constant binding energies
$E_{\mathrm{A_{1}}}$ and $E_{\mathrm{A_{2}}}$ of the two individual nuclei as obtained by the static HF initialization
\begin{equation}
V(R)=E_{\mathrm{DC}}(R)-E_{\mathrm{A_{1}}}-E_{\mathrm{A_{2}}}\ .
\label{eq:dctdhfvr}
\end{equation}
The calculated ion-ion fusion barriers incorporate all of dynamical changes in the nuclear density during the TDHF time evolution in a self-consistent manner.
As a consequence of this inclusion of dynamical effects the DC-TDHF potential is energy dependent~\cite{umar2014a}.
At high collision energies, the densities do not have time to rearrange and the results approach the frozen picture.
The features arising from the dynamic collision distinguish DC-TDHF from the fully static DCFHF approach.

\subsection{Cross sections}\label{sec:cs}
Both DCFHF and DC-TDHF provide a way to obtain one dimensional ion-ion fusion potentials, which can then be used to calculate fusion cross sections and related quantities.
However, standard implementations of barrier penetration models, i.e., with incoming wave boundary conditions (IWBC) of absorbing imaginary potentials at short distance, cannot account for the sub-barrier resonances observed experimentally~\cite{jiang2013}. 
Indeed, these boundary conditions are based on the assumption that fusion always occurs when the barrier is overcome, while in reality fusion may not happen if no corresponding states in the compound system are present. 
This is particularly critical in $^{12}$C+$^{12}$C at $E<7$~MeV due to the low level density of positive parity states in $^{24}$Mg, hindering fusion off-resonance. 
DC-TDHF predictions should then be compared with experimental cross-sections on-resonance.

Transmission probabilities $T_{l}(E_{\mathrm{c.m.}})$ are acquired by numerical integration of the two-body Schr\"odinger equation:
\begin{equation}
\left[\frac{-\hbar^2}{2\mathrm{M(R)}}\frac{d^2}{dR^2}+\frac{l(l+1)\hbar^2}{2\mathrm{M(R)}R^2} + V(R) - E\right]\psi=0 .
\label{eq:se}
\end{equation}
The IWBC method is used to calculate transmission probabilities which assumes that fusion occurs once the minimum of V(R) is reached~\cite{rawitscher1964}.
The barrier penetrability $T_l(E_{\mathrm{c.m.}})$ is then the ratio of the incoming flux at the minimum of the potential inside the barrier to the incoming flux at a large distance.
Once $T_l(E_{\mathrm{c.m.}})$ is calculated, the fusion cross sections at energies above and below the barrier are calculated as 
\begin{equation}
\sigma_f(E_{\mathrm{c.m.}})=\frac{\pi}{k_0^2}\sum_{l=0}^{\infty}(2l+1)T_l(E_{\mathrm{c.m.}}).
\end{equation}

In the case of DCFHF, the mass is a constant and equal to the reduced mass, $\mathrm{M(R)}=\mu$. 
In the case of DC-TDHF, the coordinate-dependent mass $M(R)$ can
be calculated directly from TDHF dynamics~\cite{umar2009b}.
This mass primarily influences the inner part of the barrier, leading to a somewhat broader
barrier width thus leading to further hindrance in the sub-barrier region.
Instead of solving the Schr\"odinger equation using the coordinate-dependent mass M(R), the potential can be transformed by a scale factor~\cite{umar2009b,goeke1983}
\begin{equation}
d\bar{R}=\left(\frac{\mathrm{M(R)}}{\mu}\right)^{\frac{1}{2}}dR.
\end{equation}
Upon making this transformation M(R) is replaced by the reduced mass $\mu$ in Eq.~\ref{eq:se} and the Schr\"odinger equation is solved using the modified Numerov method as it is formulated in the coupled-channel code CCFULL~\cite{hagino1999}.

\subsection{Energy density functional}

Unless otherwise stated, the Skyrme parametrization used was UNEDF1~\cite{kortelainen2012}.
To investigate the dependence on the choice of the energy density functional we have also performed calculations using the SLy4d functional~\cite{kim1997}. 
Both parametrizations were fitted without the center of mass correction, making them ideal for dynamic calculations.

These parametrizations do not include the tensor terms of the functional.
The effects of these terms on heavy-ion fusion have been recently studied~\cite{dai2014a,stevenson2016,guo2018}.
However, a comparison of the nucleus-nucleus potentials calculated with the SLy5 functional (without tensor)~\cite{chabanat1998a} and with SLy5t (including tensor)~\cite{colo2007}  showed that the effect is quite small for \textsuperscript{12}C + \textsuperscript{12}C~\cite{guo2018b}.

Furthermore, to investigate the behavior of a different density dependence than that of the one used in Skyrme fits, the QMC-I~\cite{stone2016} parametrization of the functional based on the quark-meson coupling model~\cite{guichon1988,guichon1996} was utilized for the first time in a dynamic fusion study.

\subsection{Numerical details}

All calculations were done using the three dimensional VU-TDHF3D code which contains no symmetry restrictions and implements the full energy-density functional including all time-odd terms~\cite{umar2006c}. 
The numerical box used was 72 fm in length along the collision axis and 24~fm in the other two directions.
The basis-spline collocation method used to represent derivative operators on the lattice is robust and allows for a coarser grid spacing (typically 1 fm).
However a finer grid spacing of 0.8~fm was used for the current work to ensure very precise calculations. 
When comparing barrier heights, however, the change between the coarse and fine grid spacings amounts to a difference of about $\Delta \mathrm{V}_{\mathrm{peak}}=0.5$~keV.

\section{Results}
\label{sec:res}

\begin{figure}
	\includegraphics*[width=8.6cm]{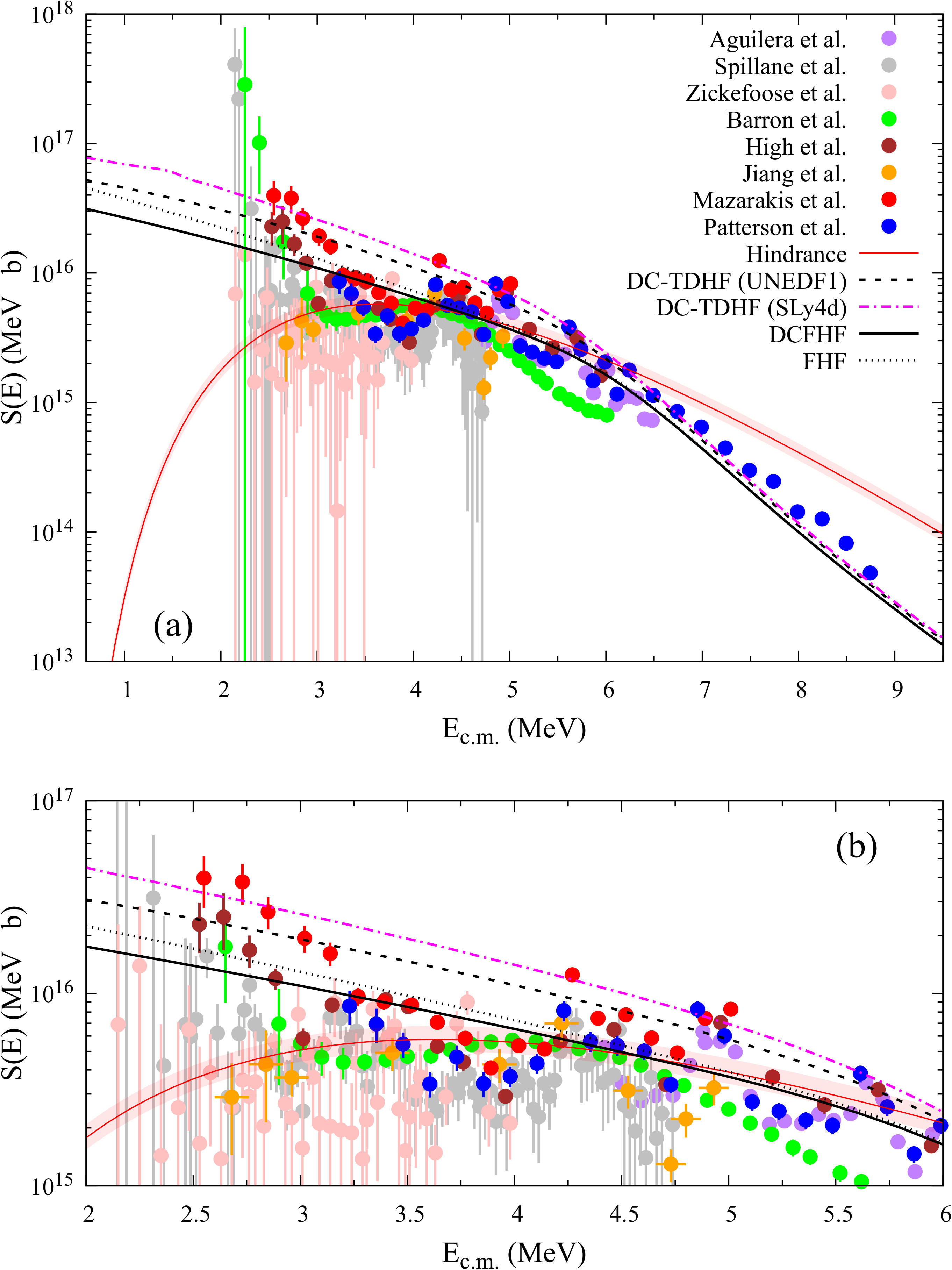}
	\caption{(Color online) Comparison of \textsuperscript{12}C+\textsuperscript{12}C fusion $S$ factors from DC-TDHF, FHF, and DCFHF using the UNEDF1 force, DC-TDHF using the SLy4d force, the hindrance model from~\cite{jiang2007}, and experimental data from~\cite{aguilera2006,spillane2007,zickefoose2011,barron-palos2006,high1977,jiang2018,mazarakis1973,patterson1969}. }
	\label{fig:sfac}
\end{figure}

In this section we present ion-ion potentials, fusion cross sections, and astrophysical $S$ factors obtained from the methods outlined in Sec.~\ref{sec:form}.

A light system such as \textsuperscript{12}C + \textsuperscript{12}C does not exhibit a strong energy dependence in the DC-TDHF potential, though in practice collision energies close to the fusion barrier are chosen to allow for maximal rearrangement.
To this end, the energy chosen for comparison to experiment was $\mathrm{E_{c.m.}}=6.03$~MeV.
Figure~\ref{fig:pots} shows a comparison of effective ion-ion potentials as calculated using the methods presented in Sec.~\ref{sec:form}.
All three methods result in an effective potential that follows the point Coulomb potential until the nuclear overlap is enough to form the barrier peak.
Both FHF and DCFHF potentials form a similar barrier top before diverging in the inner region.
For DCFHF, a pocket is formed by the pressure resulting from the Pauli exclusion principle~\cite{simenel2017}, whereas FHF decreases rapidly at small distance and has a thinner barrier width overall at sub-barrier energies.
The potential that is created from the DC-TDHF method shows that dynamic features of the collision causes the effective potential to deviate from the incoming Coulomb potential earlier that both FHF and DCFHF.
Also, DC-TDHF results in a thinner barrier than FHF and DCFHF, which is expected to enhance sub-barrier cross sections.

In Fig.~\ref{fig:dens}, a 3D contour plot resulting from the DC-TDHF calculation is shown over a pseudocolor density plot corresponding to the position of the barrier peak, $R=8.06$~fm. 
At this position, the core contours of $\rho = 0.08$~fm\textsuperscript{-3} (solid red contour surface) are well separated at the top of the barrier. 
The opaque, gray contour is drawn to show the level of the density in the overlapping region, $\rho = 0.008$~fm$^{-3}$.
The relatively low density at the barrier peak indicates that Pauli repulsion will play a small role for this system at the barrier radius.
Despite being small, this density overlap produces enough attraction to compensate the Coulomb repulsion, leading to fusion.

A more quantitative representation of the density profile at the barrier is shown in Fig.~\ref{fig:denstrace}.
Here, the density is plotted along the collision axis. 
It is interesting to note the higher density in the neck at this separation when dynamics is accounted for (DC-TDHF).
This change of neck density, induced by the dynamics, is responsible for the lowering of the barrier in DC-TDHF, as seen in Fig.~\ref{fig:pots}.
Although this change in barrier energy is relatively small, it can significantly influence tunneling probabilities at deep sub-barrier energies. 

The potentials of Fig.~\ref{fig:pots} are used to compute fusion cross sections in Fig.~\ref{fig:xsec}.
At these scales the difference does not appear to be large, though DC-TDHF seems to slightly over predicts cross sections for this system bellow the barrier.
However, as discussed in Sec.~\ref{sec:cs}, the comparison should only be made with the maximum of the cross-section on-resonance~\cite{jiang2013}, indicating a better agreement. 

A standard representation of the sub-barrier fusion cross-section is given by the astrophysical $S$ factor 
\begin{equation}
S(E)=\sigma(E)Ee^{2\pi \eta},
\end{equation}
where $E$ is the center of mass energy, $\eta=Z_1Z_2e^2/\hbar v$ is the Sommerfeld parameter, and $v$ is the relative velocity of the nuclei $v=\sqrt{2E/\mu}$ for a system of reduced mass $\mu$.
The $S$ factor is often used to analyze fusion reactions of astrophysical interest as, to some extent, it gets rid of the strong energy dependence at sub-barrier energies.

Figure~\ref{fig:sfac} shows a comparison between theoretical and experimental $S$ factors. 
Both DCFHF and DC-TDHF predict a similar trend for deep sub-barrier energies.
However neither suggest an extreme hindrance effect such as that seen in the power law extrapolation from Jiang et al.~\cite{jiang2007}.
The zoom in the low-energy region in Fig.~\ref{fig:sfac}(b) shows that DC-TDHF reproduces well the upper bound of the resonances seen in the data, except for the lowest resonance at 2.14~MeV. 
Note, however, that this resonance is not observed in all channels and is subject to experimental controversy~\cite{tang2018}.
Similar agreement with experiment was found in earlier extrapolations such as~\cite{fowler1975,gasques2005}, though it should be reiterated that none of the ion-ion potentials presented in this work were fit to experimental data.
Note also the role of the Pauli repulsion which can be seen by comparing FHF to DCFHF~\cite{simenel2017}. 
We see that the additional fusion hindrance at sub-barrier energies due to Pauli repulsion is largely negligible at experimental energies.

\begin{figure}
	\includegraphics*[width=8.6cm]{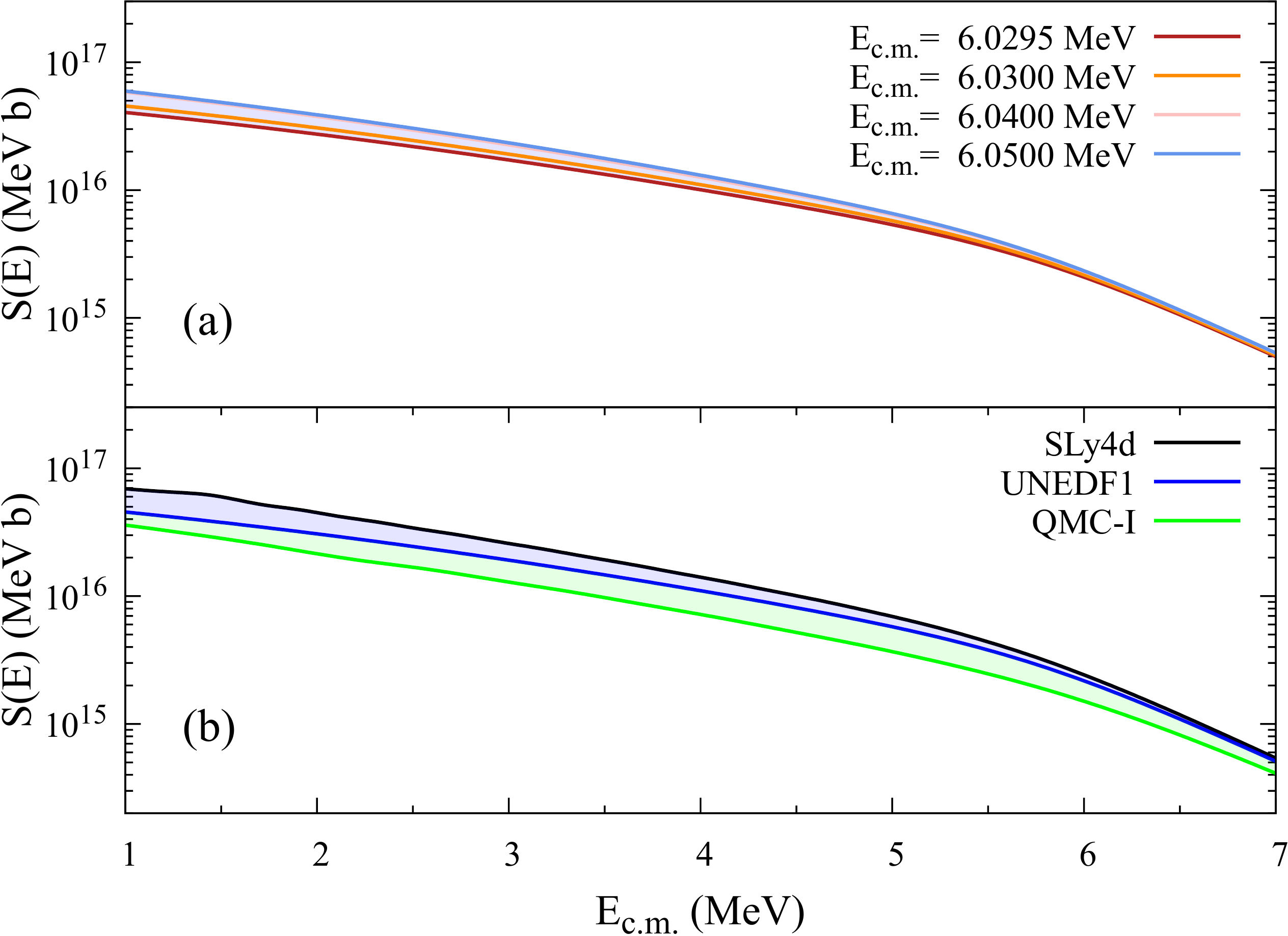}
	\caption{(Color online) Comparison of \textsuperscript{12}C+\textsuperscript{12}C fusion $S$ factors from DC-TDHF at (a) differing TDHF energies and (b) using different functionals.}
	\label{fig:sfaccomp}
\end{figure}

As mentioned in the introduction, there is no guarantee that the dynamical effects on the potential extracted from a TDHF calculation at an energy near the barrier are the same as in deep sub-barrier energies. 
To test the validity of this approximation,  DC-TDHF calculations have been performed at different TDHF energies, as shown in Fig.~\ref{fig:sfaccomp}(a).
When examining the behavior of the $S$ factors at deep sub-barrier energies, differences in fusion cross sections are greatly magnified.
Nevertheless, the difference between the DC-TDHF predictions remains small, indicating that the results are only slightly affected by the energy dependence of the potential. 
The bulk of this effect arises from the increasing and sharpening peaks of the coordinate dependent mass $M(R)$ at the location of the barrier at lower energies, though as mentioned before, density rearrangement also plays a role. 

Finally, let us investigate the sensitivity to the energy density functional. 
As such, we have plotted the comparison using the resultant $S$ factors in part (b) of Figure~\ref{fig:sfaccomp} which shows the astrophysical $S$ factor for three different functionals: the SLy4d \cite{kim1997} and UNEDF1 \cite{kortelainen2012} Skyrme functionals and the QMC-I functional which has a different density dependence than the Skyrme one. 
The main conclusion is that the trends are very similar for these three functionals. In particular, none of them predict a maximum in the $S$ factor. 
Quantitatively, the QMC results are lower throughout the sub-barrier energy region.
Nevertheless, the variations in the experimental data does not allow for the unambiguous identification of what functional best reproduces experimental results.

%

\section{Conclusion}
\label{sec:conclusion}
A precise study of \textsuperscript{12}C+\textsuperscript{12}C fusion has been performed using multiple parameter-free, microscopic approaches. 
Various energy density functionals were also explored, including the first application to heavy-ion fusion with an energy-density functional based on the quark-meson coupling model.
A comparison between FHF and DCFHF approaches shows that Pauli repulsion only plays a minor role in this system, and is not sufficient to induce a maximum in the astrophysical $S$ factor. 
Dynamical effects described in the DC-TDHF approach exhibit the same trend as static calculations, with slightly higher sub-barrier fusion cross-sections due to a narrowing of the potential barrier. 
DC-TDHF predictions are in good agreement with the maximum fusion cross-sections on-resonances. 
The different functionals all lead to the same trends in the astrophysical $S$ factor, with slightly smaller values for the QMC-I functional. 

Future studies of other light nuclei should be performed using both DCFHF and DC-TDHF as the effect of dynamic processes may play a larger role in other reactions, e.g. transfer effects in asymmetric systems such as \textsuperscript{13}C+\textsuperscript{12}C. 
The separation of static and dynamic effects is an interesting endeavor in itself which may hint at what drives (and hinders) fusion in both light and heavy nuclei.

Finally, to address the limitations inherent in the methods pursued here and to further understand the fusion process for systems like \textsuperscript{12}C+\textsuperscript{12}C, additional techniques for studying sub-barrier fusion should be explored.
One such improvement would be to pursue a fully microscopic description of many-body tunneling, e.g., following Refs.~\cite{levit1980c,reinhardt1980}, 
 to avoid the reduction of the problem to the two-body case as done here.
Such a method would be of great use beyond light systems or fusion alone, opening the door for a fully microscopic mean field description of fission and fusion.

\begin{acknowledgments}
We thank S. Courtin for useful discussions and for clarifying experimental questions. 
This work has been supported by the
Australian Research Council Grants No. DP180100497 and DP190100256,
and by the U.S. Department of Energy under grant No.
DE-SC0013847 with Vanderbilt University.
Part of this research was undertaken with the assistance of resources
from the National Computational Infrastructure (NCI), which is supported by the Australian Government.
\end{acknowledgments}

\bibliography{VU_bibtex_master}

\end{document}